# Network Coding Implementation Details: A Guidance Document

Somayeh Kafaie, Yuanzhu Peter Chen[*], Octavia A. Dobre, Mohamed Hossam Ahmed
Faculty of Engineering and Applied Science, [*]Department of Computer Science
Memorial University of Newfoundland
St. John's, NL, A1B3X5, Canada

*Abstract*- In recent years, network coding has become one of the most interesting fields and has attracted considerable attention from both industry and academia. The idea of network coding is based on the concept of allowing intermediate nodes to encode and combine incoming packets instead of only copy and forward them. This approach, by augmenting the multicast and broadcast efficiency of multi-hop wireless networks, increases the capacity of the network and improves its throughput and robustness. While a wide variety of papers described applications of network coding in different types of networks such as delay tolerant networks, peer to peer networks and wireless sensor networks, the detailed practical implementation of network coding has not been noted in most papers. Since applying network coding in real scenarios requires an acceptable understanding of mathematics and algebra, especially linear equations, reduced row echelon matrices, field and its operations, this paper provides a comprehensive guidance for the implementation of almost all required concepts in network coding. The paper explains the implementation details of network coding in real scenarios and describes the effect of the field size on network coding.

## I. Introduction

The idea of network coding was introduced in 2000 by Ahlswede [1], and has attracted a lot of attention from research community. The principle behind network coding is to allow intermediate nodes to combine received packets in one packet and forward it. By applying this idea the transmission capacity of the network increases and even it is possible to reach the full theoretical capacity of the network [2].

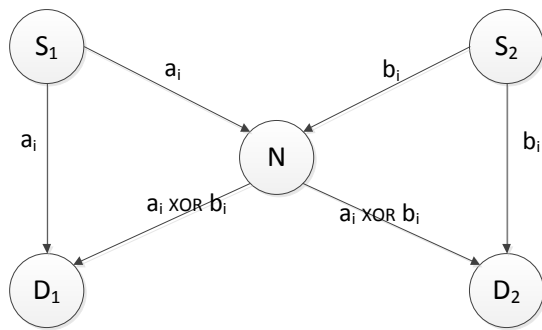

Figure 1. Butterfly model showing how network coding improves throughput.

The famous butterfly example [3], depicted in Figure 1, illustrates the concept in more details. As is shown in this figure, $S_1$ and $S_2$ send packets $a_i$ and $b_i$ respectively to both $D_1$ and $D_2$. Without network coding the intermediate node $N$ at each time unit is able to deliver either $a_i$ or $b_i$. So, the middle link becomes a bottleneck and decreases the throughput of the network. On the other hand, when network coding is applied, the intermediate node $N$ combines (encodes) $a_i$ and $b_i$ (e.g., by XOR), enables both $D_1$ and $D_2$ to receive two new packets (i.e., from either $S_1$ or $S_2$, and $N$) at each time unit, improves the throughput. Note that $D_1$ (resp. $D_2$) can obtain $b_i$ (resp. $a_i$) by XOR-ing the packet received from $N$ ($a_i$ XOR $b_i$) and the one received directly from the source, $a_i$ (resp. $b_i$).

In last decade, a lot of research has been conducted in this area, and network coding has found considerable application in different types of networks such as wireless networks, cognitive radio networks, peer to peer networks, and ad-hoc sensor networks [2, 4]. Although the idea of network coding seems simple and straightforward, its implementation in real scenarios needs an adequate level of knowledge about some mathematical and algebraic concepts especially in finite fields and systems of linear equations.

This paper explains required information to realize network coding, and provides guidance for researchers who plan to work in network coding area. In what follows, Section II presents the definition of finite fields and their application in network coding. Section III provides an overall view of implementation of network coding, and explains how it works. Section IV describes the implementation of network coding for the smallest field size (a simple case). Section V discusses some challenges introduced by increasing the field size and extends the implementation. Finally, Section VI concludes the paper and presents some future work.

## II. Finite Field and Network Coding

### A. Finite Field Definition

In algebra a field is a set whose members are from communicative group under addition and multiplication (except for zero in multiplication). In fact, a field could be interpreted as an abstraction of other numerical systems like real numbers with the same basic characteristics. A finite field or Galois field (GF) is a field with finite number of elements.

### B. Finite Field in Network Coding

The specific properties of finite fields are beneficial in network coding. Network coding is about processing some packets consisting of binary digits which can be assumed as the members of a binary finite field. So, finite fields enhance the process of encoding and decoding of the packets. For example, a packet 1101001 with 6 bits can be treated as a member of GF($2^6$).

The field size shows the range of the members of the field. For example, the field GF($2^8$) with size equal to $2^8$ has 256 integer members from 0 to 255(i.e., a byte or 8 bits with binary representation of 256 integer values). In network coding, to encode a packet, it is first divided to symbols with $s$ bits length, while the field size is $2^s$. So, each symbol M is the binary representation of an integer value within the range of the field (i.e., M ∈ [0, $2^s$-1]). It is notable that the encoding and decoding process are accomplished on the symbols of the packets.

## III. HOW NETWORK CODING WORKS?

The basic idea of network coding is to combine the packets in each node instead of just forwarding them. Therefore, when more than one packet are available to send in the source node or each intermediate node, the packets are combined and encoded packets are sent. Then it is the responsibility of the destination to decode the received packets and extract the original ones. Note that the intermediate nodes just encode the received packets and forward them; they do not need to decode the packets.

### A. Encoding

In the encoding process, whenever a node, either the source node or any intermediate node, has to forward a packet, it checks its waiting queue and if more than one packet are available, it generates a linear combination of them as an encoded packet and sends it. So, the node needs to select a vector of coefficients which represents the weight of each packet in the encoded packet. The most popular way to select these coefficients is to select them from a uniformly random distribution over the finite field. Although it is probable that some of the random combinations are not independent, especially for smaller field sizes, research shows that this probability is too low and insignificant [4].

### B. Encoding Vector and Information Vector

For the destination to be able to recover the original packets, it has to know how they have been combined together. Therefore, each node should send an *encoding vector*, which represents the weight of different original packets in the encoded packet, along with the encoded data which is called *information vector*. General speaking, each forwarded encoded packet indicates an equation like

$$e_1 X_1 + e_2 X_2 + \cdots + e_n X_n = Y, \qquad (1)$$

where $n$ is the number of original packets, $X_i$ is $i$th original packet, $e_i$ is $i$th element of the encoding vector corresponding to $X_i$, and $Y$ is calculated encoded data or information vector. Since we noted earlier that all operations are done on the symbols of packets instead of the whole packet, the previous equation can be rewritten in symbol level as

$$e_1 \prod_{j=1}^{m} X_1^j + e_2 \prod_{j=1}^{m} X_2^j + \cdots + e_n \prod_{j=1}^{m} X_n^j = \sum_{i=1}^{n} e_i \prod_{j=1}^{m} X_i^j = \prod_{j=1}^{m} Y^j, \qquad (2)$$

where Π shows the concatenation operation between symbols of a packet and $m$ is the number of symbols in each packet which is calculated by $packet\_size/s$ for GF($2^s$).

### C. Encoding in Intermediate Nodes

It is clear that for the source node the encoding process is summarized in choosing random coefficients and then apply them in Equation (2) to calculate the encoded packet Y (i.e., information vector). In this case the encoding vector is equal to the randomly selected coefficients. However in intermediate nodes the situation is a little different, since they do not have the original packets and also the number of packets in their waiting queue is different from the number of the original packets. So the problem is how to combine these received encoded packets in a way that the elements of the calculated encoding vector, after combination (re-encoding), correspond to the original packets.

Let say the waiting queue of the intermediate node $A$ contains $r$ received encoded packets with information vectors $X'_1, \ldots, X'_r$ and encoding vectors $E'_1, \ldots, E'_r$. To generate a new encoded packet (i.e., information vector) based on the received ones, node $A$ picks $r$ coefficients $w_1, \ldots, w_r$ which represent the weight of the received packets in the new encoded packet Y, and calculate Y by

$$Y = w_1 X'_1 + w_2 X'_2 + \cdots + w_r X'_r, \qquad (3)$$

where $w_i$ shows the coefficient corresponding to $i$th received packet ($X'_i$), and like Equation (2) all calculations are in symbol level instead of packet level.

For encoded packets generated by intermediate nodes the encoding vector is not equal to the randomly selected coefficients, because it should be computed with respect to the original packets. The encoding vector $E = (e_1, \ldots, e_n)$ is calculated in Equation (4) by multiplication of coefficient vector ($w_1, \ldots, w_r$) in Matrix $F_{r \times n}$, where $i$th row of the matrix is $E'_i$. In the other word, each element of the encoding vector is calculated by $e_i = \sum_{j=1}^{r} w_j \times E'_j(i)$ [4].

$$E = W_{1 \times r} \times F_{r \times n}, \qquad (4)$$

### D. Decoding

The destination receives some encoded packets $Y_i$ in the form represented in Equation (1) which create an n-variable system of linear equations, and decoding the packets is equivalent to finding a solution for this system. To solve the system and find $X_i$s (i.e., original packets), the destination stores the received encoding and information vectors in a matrix and keep it in reduced row echelon form by using Gaussian elimination algorithm.

A matrix is in reduced row echelon form if the leading coefficient of each row is 1, it is the only nonzero in its column, and also the position of the leading coefficient of each row is always to the right of above rows [5]. Whenever the destination receives a new encoded packet, it adds this packet to the matrix and applies Gaussian elimination algorithm to keep it in reduced row echelon form. If the new arrived packet increases the rank of the matrix, it is called *innovative*, and stored in the matrix. Otherwise, if it is reduced to a row of 0s, it is ignored. Figure 2 presents an overview of network coding flowchart.

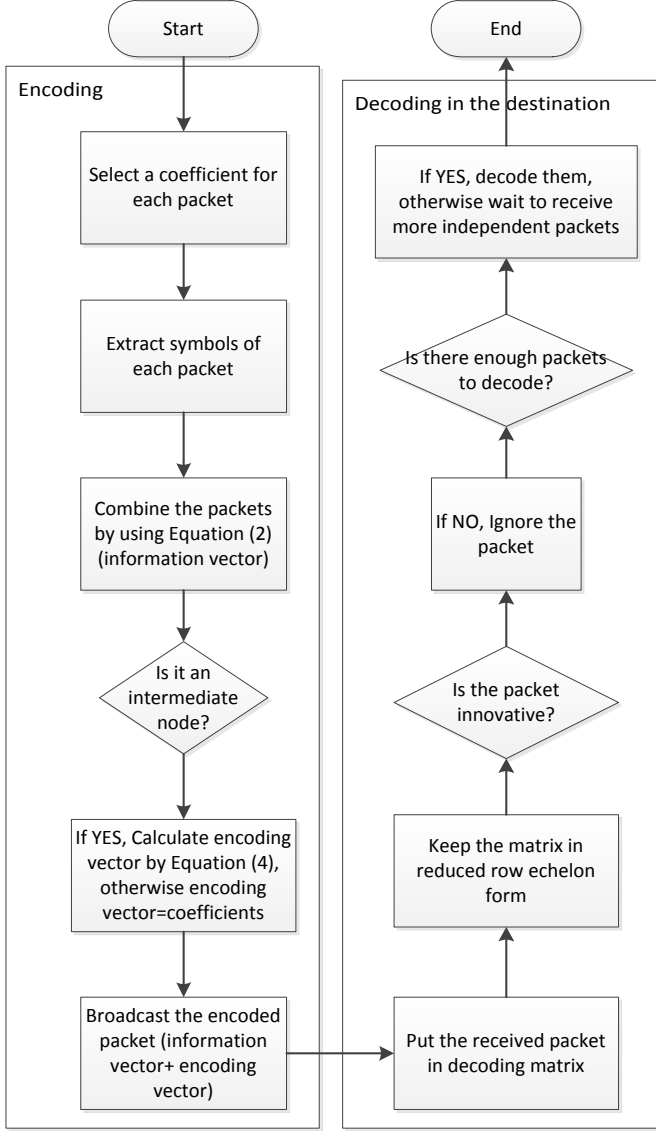

Figure 2. Network coding flowchart.

## IV. SIMPLE CASE OF $GF(2^1)$

In this section linear network coding is described for simple case of $GF(2^1)$ (i.e., $s = 1$). Then in the next section we will extend the work for larger field sizes. For the field size of $2^1$ the symbol size is equal to 1. It means that each bit of the packet represents a symbol, and a small data type like boolean can be used for symbols. Since the coefficients are picked randomly over the field range, for s=1 each coefficient can be either 0 or 1. So in this case, multiplication of each original packet in its corresponding element of the encoding vector (i.e., either 0 or 1) in Equation (1) and (3) simply shows whether the original packet has participated in the combination or not. For example for encoding vector (0, 1, 0, 0, 1) the calculated information vector is $Y = X_2 + X_5$ which shows the second and the fifth original packets have been combined together.

As it is noted before all operations for encoding and decoding are in the symbol level. It means that to multiply a coefficient by a packet, we multiply it by its symbols one by one, and to add two packets we add their corresponding symbols together. If the packets to be added (combined) do not have the same size, enough zeros are attached to the end of the smaller one. Since symbols have been defined over the finite filed set, the result of multiplication and addition must be in this set. For s=1, since coefficients are either 0 or 1, we assure that the multiplication of the symbols and coefficients is within the range of the finite field, but for addition *bitwise XOR* is used to keep the symbols in this range.

Figure 3 shows how linear network coding works in a simple scenario with 3 nodes and the finite field $GF(2^1)$. Let say the source node *S* plans to send 4 packets $P_1, \ldots, P_4$ to the destination *D*. It combines them together by selecting 4 different coefficient sets or encoding vectors, generates $Q_1, \ldots, Q_4$, and broadcasts them. The intermediate node *A* receives 3 packets $Q_1, Q_2, Q_3$, selects a coefficient set $W = (w_1, w_2, w_3)$ to combine these 3 packets, generates $Q_{A1}$, calculate its encoding vector, and forward it.

When D receives the packets it decodes them and finds the original packets $P_1, \ldots, P_4$ by solving the provided system of linear equations. Although it must receive at least 4 *independent* encoded packets (i.e., as many as the number of original packets) to be able to decode all packets, it can decode a packet if its corresponding row in reduced row echelon form has only one nonzero (i.e., 1) element. In this figure, for example, $Q_4$ is enough for D to decode $P_4$.

## V. IMPLEMENTATION FOR LARGER FIELDS

When the finite field size increases, let say $2^s$ and $s > 1$, the symbol length is not one bit anymore, and increases. Also the coefficients and the elements of the encoding vector are picked from a larger set of [0, $2^s$-1]. So, the probability of generating independent encoded packets is higher. On the other hand, increasing the field size makes encoding and decoding operations more challenging, as an efficient data type should be selected for symbols and coefficients, finite field operations such as multiplication, subtraction, and inversion have to be defined, and also more computational operations are performed to keep the decoding matrix in reduced row echelon form.

While for $GF(2^1)$ the symbols can be stored in boolean data type, for larger fields larger data types are required. Since field sizes larger than 32 is not practical and are hardly used, after conducting some experiments in Java, we decided to choose data type *int* to store symbols and coefficients. It should be noted that *int* in Java takes 32 bits.

### A. Multiplication

For any $GF(2^s)$, the subtraction can be implemented like addition by *bitwise XOR*. So, if $a + b = a \oplus b = c$, then $c - a = c \oplus a = b$. To multiply two numbers in the finite field, it is important to make sure that the result is within the finite field. To obtain multiplication, the binary representation of each number like $a_n a_{n-1} \ldots a_1 a_0$ is mapped to the polynomial $a_n Z^n + \cdots + a_1 Z + a_0$. In addition, an irreducible polynomial of degree $s$ over $GF(2^s)$ is picked modulo which all operations on $GF(2^s)$ are performed. A list of such irreducible polynomials for different field sizes has been introduced by Seroussi [6]. For example, the most common irreducible polynomial for $GF(2^8)$ is $Z^8 + Z^4 + Z^3 + Z + 1$ with 100011011 binary representation.

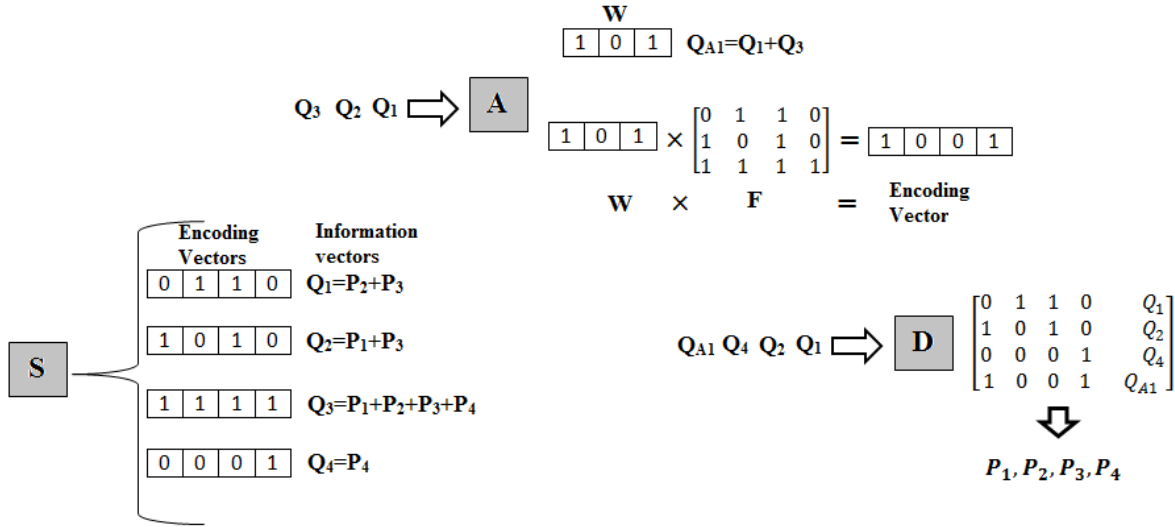

Figure 3. Encoding and decoding process.

When the result of the usual product of two numbers is larger than $2^s$-$1$ (i.e., the degree of its polynomial is larger than $s$-$1$), this product modulo selected irreducible polynomial represents the result of multiplication. Let calculate the product of 10 and 41 in $GF(2^8)$, as an example. Their binary representations are 00001010 and 00101001, respectively. We multiply their polynomial form together as follows:
$10 \times 41 = (z^3 + z)(z^5 + z^3 + 1) = z^8 + z^6 + z^3 + z^6 + z^4 + z = z^8 + z^4 + z^3 + z = 100011010$

Since XOR is used for addition, two $x^6$s are eliminated. Now, because the degree of the result is larger than 7 (i.e., $s$-$1$), the result of the product modulo 100011011 (i.e., our irreducible polynomial) presents the final multiplication result which is equal to 1 (i.e., $100011010 \bmod 100011011 = 1$). So, 10 times 41 is equal to 1 in $GF(2^8)$. Note that to calculate the remainder the binary division is used in which XOR has been applied instead of the arithmetic subtraction.

A simple algorithm to implement the multiplication [7] is depicted in Figure 4. It should be noted that in this algorithm, which just uses shifting and XOR-ing, the highest term of the irreducible polynomial should be eliminated. For example, 100011011 which is the binary representation of the irreducible polynomial for $GF(2^8)$, is converted to 11011. Our experiments show that while for smaller field sizes (e.g., $s < 8$) providing a multiplication table is more efficient, for larger field sizes, due to memory limits, multiplication is done on demand and on the fly.

B. *Inversion*

To keep decoding matrix in reduced row echelon form four operations are required as addition, subtraction, multiplication and inversion. A multiplicative inverse or simply inverse of the number z, denoted by 1/z or $z^{-1}$, is a number whose multiplication by t is equal to multiplication identity, 1. So, in $GF(2^8)$, 41 is considered as the inverse of 10, because $10 \times 41 = 1$. If the multiplication table is available, the inverse of each number can be calculated simply by searching the corresponding row of the number to find the column index of a cell containing 1. Otherwise, the inverse has to be calculated by one of available algorithms. Figure 5 demonstrates an inversion method based on Itoh-Tsujii addition chain [8, 9].

```
Inputs: a, b,
q (the irreducible polynomial with highest term
eliminated),
s (as multiplication is over GF(2^s))
Output: P = a × b
multiplication(a, b)
{
  For (i=0; i<s; i++)
  {
    If (the rightmost bit of b is 1)
      P = P ⊕ a; // P XOR a
    If (the leftmost bit of a is 1)
      carry = 1;
    a = a << 1; //left shift
    If (carry == 1)
      a = a ⊕ q;
    b = b >>> 1; //unsigned right shift
  }
  return P; //P = a × b
}
```

Figure 4. Multiplication method.

```
Inputs: a = $A_{s-1}A_{s-2}...A_1A_0$  ($a \neq 0$),
q (the irreducible polynomial with highest term
eliminated),
s (as multiplication is over GF($2^s$)),
$h_r h_{r-1} ... h_1 h_0$ is the binary representation of s-1
($h_r \neq 0$)
Output: c = a$^{-1}$ //the inverse of a
getInverse(a)
{
  c = a;
  k = 1;
  For (i=r-1; i>=0; i--)
  {
     b = c;
     For (j=0; j<k; j++)
        b = multiplication(b , b); //b × b;
     c = multiplication(c , b); //c × b
     k = 2 × k;
     If ($h_i == 1$)
     {
        c = multiplication(c, c);
        c = multiplication(c, a); //c × c × a
        k = k + 1;
     }
  }
  c = multiplication(c, c); //c × c
  return c; //c = a$^{-1}$
}
```

Figure 5. Inversion method.

## VI. CONCLUSION AND FUTURE WORK

Network coding is an amazing idea which enables the nodes to decode the incoming packets before forwarding them, so increases the capacity of the network and improves throughput and robustness. In last decade, network coding has become one of the most important and attractive topics and a lot of research has been done in this area. Although implementation of network coding necessitates adequate understanding of some mathematical and algebraic concepts like finite field operations, system of linear equations, and reduced row echelon form, such primary requirements hardly have been noted in network coding publications, making hard its implementation for novice researchers of this area. This paper provided a guidance document, and explained the basic algebra and mathematic required to realize network coding.

This paper described implementation details of network coding with regard to the field size and a fixed number of original packets, and considering the effect of unknown number of original packets is left as future work. When the number of original packets is not specified or there is memory limitation, packets are grouped together in *generations*, and only packets in the same *generation* are allowed to be combined with each other. This idea introduces new implementation challenges such as how to define generation size, and how to map each packet to one specific generation which can be considered as future work.

ACKNOWLEDGMENT

This work was supported in part by the Natural Sciences and Engineering Research Council (NSERC) of Canada (Discovery Grant 327667-2010, 327285-2013 and 293287-2009).

REFERENCES

[1] R. Ahlswede, N. Cai, S. R. Li and R. W. Yeung, "Network Information Flow" *IEEE TRANSACTIONS ON INFORMATION THEORY,* vol. 46, pp. 1204-1216, 2000.
[2] M. Sanna and E. Izquierdo, "A Survey of Linear Network Coding and Network Error Correction Code Constructions and Algorithms," *International Journal of Digital Multimedia Broadcasting,* vol. 2011, pp. 1-12, 2011.
[3] S. Katti, H. Rahul, W. Hu, D. Katabi, M. Medard and J. Crowcoft, "XORs in The Air: Practical Wireless Network Coding " presented at the SIGCOMM'06, Pisa, Italy, 2006.
[4] C. Fragouli, J. L. Boudec and J. Widmer, "Network Coding: An Instant Primer," *ACM SIGCOMM Computer Communication Review,* vol. 36, pp. 63-68, 2006.
[5] S. Leon, *Linear Algebra with Applications*: Pearson, 2009.
[6] G. Seroussi, "Table of Low-Weight Binary Irreducible Polynomials," Hewlett-Packard CompanyAugust 1998.
[7] S. Trenholme. 14-10-2013. AES' Galois field. Available: http://www.samiam.org/galois.html
[8] A. Mahboob and N. Ikram, "Faster Polynomial Basis Finite Field Squaring and Inversion for GF(2m) With Cryptographic Software Application," presented at the International Symposium on Biometrics and Security Technologies (ISBAST '08) Islamabad, Pakistan, 2008.
[9] A. Abdulah Zadeh, "Division and Inversion Over Finite Fields," in *Cryptography and Security in Computing*, J. Sen, Ed., ed: InTech, 2012, pp. 117-129.